\documentclass[12pt,thmsa]{article}

\usepackage{amsfonts}
\usepackage{graphicx}
\usepackage{epsfig}
\usepackage{graphics}

\makeatletter
\newcommand{\row}[1]%
{\mathord{\buildrel{\lower3pt%
\hbox{$\scriptscriptstyle\rightarrow$}}\over #1}}
\newcommand{\col}[1]{{#1^{\raisebox{2pt}[\height]%
{$\scriptstyle\downarrow$}}}}
\newcommand{\dyadic}[1]{\mathord{\dyadic@rrow{#1}}}
\newcommand{\dyadic@rrow}[1]{
\begin{picture}(12,12)(-1,0)
\put(-3,12){\makebox(0,0)[t]{$\scriptscriptstyle\downarrow$}}
\put(-3,13){\makebox(0,0)[l]{$\scriptscriptstyle\longrightarrow$}}
\put(5,0){\makebox(0,0)[b]{$#1$}}
\end{picture}
}

\newcommand{\ket}[1]{\bigl| #1 \bigr\rangle}
\newcommand{\expect}[1]{\left\langle #1 \right\rangle}

\newcommand{\JMO}[3]{J. Mod.\ Opt.\ \textbf{#1}, #2 (#3)}

\input{tcilatex}
\begin{document}
\begin{center}
{\Large Quantum dense coding over Bloch channels}

N. Metwally \\[0pt]

Math. Dept., Faculty of Science, South Valley University, Aswan, Egypt. \\[0pt]
E.mail: Nmetwally$@$gmail.com
\end{center}
 \begin{abstract}

Dynamics of coded information over Bloch channels is investigated
for different values of the channel's parameters. We show that,
the suppressing of the travelling coded information over Bloch
channel can be increased by decreasing the equilibrium absolute
value of information carrier and consequently decreasing the
distilled information by eavesdropper.
 The amount of decoded information
can be improved by increasing the equilibrium values of the two
qubits and decreasing  the ratio between longitudinal and
transverse relaxation times. The robustness of coded information
in maximum and partial entangled states is  discussed. It is shown
that  the maximum entangled states are more robust than the partial
entangled state over this type of channels.

{\bf Keywords:} Dense coding,  Channels, Local and non-local information.\\

\end{abstract}

\section{Introduction}

Decoherence  represents  the most difficult obstacles in  quantum
information processing. This unavoidable  phenomena can be seen in
different pictures such as, the undesirable interactions between
 systems and their surroundings \cite{Ekert}, device
imperfections \cite{Gero,Sug}, decay due to spontaneous, emission
and noisy channel \cite{MFS}. These  interactions corrupt the
information stored in the system and  consequently cause errors in
the transferred information. Therefore, investigating the dynamics
of information  in the presence of decoherence is one of the most
important tasks in quantum computation and information.

 Quantum coding is  one  of techniques that   has been used to transfer
information between two users \cite{Ben}. To achieve quantum
coding protocol with high efficiently, one needs  maximum
entangled state and ideal channels. There are  some protocols
which have been presented different treatments of quantum   coding
over noseless channels theoretically \cite{Bruss, Liu} and
experimentally \cite{Matle}. In real word, it is very difficulty
to keep  systems  which are used for quantum coding isolate.
Therefore, it is important to introduce quantum  coding  protocols
over noisy channels. Recently, Shadman and et.al., \cite{Shad}
have investigated super dense coding over  noise, where they
consider the case of Pauli channels in arbitrary dimension and
derive the super dense coding capacity.

 In the present work,  we introduce different type of quantum
 channels called Bloch channels. The decoherence effect of these
 channels on the entanglement and information has been  investigated by
 Ban and et. al. \cite{Ban-1, Ban-2}. They  considered
 one qubit passing through the Bloch channel. Metwally \cite{Nasser09} have investigated the effect of
the Bloch cannel on the fidelity of the teleported state, where
the two qubits are pass through the channel. This motivated us to
investigate  the effect of the Bloch channels on the dynamics of
coded information. Also, in this context the behavior of the local
and non-local information is studied.

 The paper is organized as follows: In Sec.$2$, we examine  the evolution  of a general
 two-qubits
state passes through Bloch channel. The quantum  dense coding is
discussed in Sec.3. The dyanmics of the local and non-local
information  is investigated in Sec.4. Finally, we discuss our
results  in Sec. 5.

\section{ Model and its solution}
 The characterization
of the 2-qubit states produced by some sources requires
experimental determination of 15 real parameters. Each qubit is
determined by $3$ parameters, representing the Bloch vectors, and
the other $9$ parameters represent the correlation tensor. Analogs
of Pauli's spin operators are used for the description of the
individual qubits; the set  $\sigma_{1x},\sigma_{1y}, \sigma_{1z}$
for  the first qubit and $\sigma_{2x},\sigma_{2y}, \sigma_{2z}$
for  the second  qubit. Any two qubits state is described by
\cite{Gunter,NM, wang},

\begin{equation}
\rho_{ab}=\frac{1}{4}(1+\row{S}\cdot\col{\sigma_1}+\row{R}\cdot\col{\sigma_2}+\row\sigma_1
\cdot\dyadic{Q}\cdot\col{\sigma_2}),
\end{equation}
where $\row\sigma_1$ and $\row\sigma_2$ are the Pauli's spin
vectors of the first and the second qubit respectively. The
statistical operators for the individual qubits are specified by
their Bloch vectors, $\row{A}=\expect{\row\sigma_1}$ and
$\row{R}=\expect{\row\sigma_2}$. The cross dyadic, $\dyadic{Q}$ is
represented  by a $3\times 3$ matrix, it describes the correlation
between the first qubit,
$\rho_a=tr_{b}\{\rho_{ab}\}=\frac{1}{2}(1+\row{S}\cdot\col{\sigma_1})$
and the second qubit,
$\rho_b=tr_{a}\{\rho_{ab}\}=\frac{1}{2}(1+\row{R}\cdot\col{\sigma_2})$.
The Bloch vectors and the cross dyadic are given by
\begin{eqnarray}
\row{S}&=&(s_x,s_y,s_z) ,\quad \row{R}=(r_x,r_y,r_z),\quad
\mbox{and}\quad \dyadic{Q}= \left(
\begin{array}{ccc}
q_{11}&q_{12}&q_{13}\\
q_{21}&q_{22}&q_{23}\\
q_{31}&q_{32}&q_{33}
\end{array}
\right).
\end{eqnarray}
\\
 Let us consider that each qubit is forced to pass through
Bloch channel. This type of  channels is  defined by the Bloch
equations \cite{Ban-1}, for the first qubit,
\begin{eqnarray}\label{ch1}
\frac{d}{dt}\expect{\sigma_{1x}}_t&=&-\frac{1}{T_{2a}}\expect{\sigma_{1x}}_t,~
\frac{d}{dt}\expect{\sigma_{1y}}_t=-\frac{1}{T_{2a}}\expect{\sigma_{1y}}_t,
\nonumber\\
\frac{d}{dt}\expect{\sigma_{1z}}_t&=&-\frac{1}{T_{1a}}(\expect{\sigma_{1z}}_t-\expect{\sigma_{1z}}_{eq}),
\end{eqnarray}
while for the second qubit, they are given by
\begin{eqnarray}\label{ch2}
\frac{d}{dt}\expect{\sigma_{2x}}_t&=&-\frac{1}{T_{2b}}\expect{\sigma_{2x}}_t,~
\frac{d}{dt}\expect{\sigma_{2y}}_t=-\frac{1}{T_{2b}}\expect{\sigma_{2y}}_t,
\nonumber\\
\frac{d}{dt}\expect{\sigma_{2z}}_t&=&-\frac{1}{T_{1b}}(\expect{\sigma_{2z}}_t-\expect{\sigma_{2z}}_{eq}),
\end{eqnarray}
where $T_{1i}$ and $T_{2i}$,~ $i=a,b$  are the longitudinal and
transverse relaxation times for  the first and the second qubit, and
$\expect{\sigma_{1z}}_{eq}$,~ $\expect{\sigma_{2z}}_{eq}$ are the
equilibrium values of $\expect{\sigma_{1z}}_t$ and
$\expect{\sigma_{2z}}_t$ respectively. Now, we  assume that the two
qubits pass through the channels (\ref{ch1}), and (\ref{ch2}). Then
the output state  is given by \cite{Nasser09},
\begin{equation}\label{out}
\rho_q(t)=\frac{1}{4}\Bigl(1+\row{S(t)}\cdot\col{\sigma_1}+\row{R(t)}\cdot\col{\sigma_2}
+\row{\sigma_1}\cdot{\dyadic{Q}(t)}\cdot\col{\sigma_2}\Bigr),
\end{equation} where,
\begin{eqnarray}\label{out1}
\row{S(t)}&=&(s_x\beta_1,~-\beta_1 s_y,~\gamma_1
s_z+(1-\gamma_1)\expect{\sigma_{1z}}_{eq}),
\nonumber\\
\row{R}(t)&=&(\beta_2 r_x,~-\beta_2 r_y,~\gamma_2
r_z+(1-\gamma_2)\expect{\sigma_{2z}}_{eq}),
 \nonumber\\
 Q_{xx}(t)&=&\beta_1\beta_2 q_{11},\quad \tilde
Q_{xy}(t)=-q_{12}\beta_1\beta_2,
 \nonumber\\
Q_{xz}(t)&=&\beta_1\gamma_2q_{13}+\beta_1(1-\gamma_2)\expect{\sigma_{z1}}_{eq}
s_x,
\nonumber\\
 Q_{yx}(t)&=&-q_{12}\beta_1\beta_2,~
Q_{yy}(t)=q_{22}\beta_1\beta_2,
 \nonumber\\
 Q_{yz}(t)&=&-\beta_1\gamma_2q_{23}-\beta_1(1-\gamma_2)\expect{\sigma_{2z}}_{eq}s_y,
\nonumber\\
 Q_{zx}(t)&=&\beta_2\gamma_1q_{31}+\beta_2(1-\gamma_1)\expect{\sigma_{z1}}_{eq}r_x,
 \nonumber\\
 Q_{zy}(t)&=&-\beta_2\gamma_1q_{32}-\beta_2(1-\gamma_1)\expect{\sigma_{1z}}_{eq}r_y,
\nonumber\\
Q_{zz}(t)&=&\gamma_1\gamma_2q_{33}+(1-\gamma_1)(1-\gamma_2)\expect{\sigma_{z1}}_{eq}\expect{\sigma_{2z}}_{eq}
\nonumber\\
&&+\gamma_1(1-\gamma_2)\expect{\sigma_{2z}}_{eq} s_3
+\gamma_2(1-\gamma_1)\expect{\sigma_{2z}}_{eq}r_z,
\end{eqnarray}
and $\gamma_i=exp\{-\frac{t}{T_{1i}}\}, ~
\beta_i=exp\{-\frac{t}{T_{2i}}\},~\mbox{and} ~i=a,b$.

Equation (\ref{out}), represents the time evaluation of any two
qubits state passes through the Bloch channels Eqs.(\ref{ch1}) and
(\ref{ch2}). Assume that the users, Alice and Bob share one
maximum entangled state of Bell's states, $\ket{\psi^{\pm}}$ or
$\ket{\phi^{\pm}}$. The dynamics of these states can be obtained
from (\ref{out}) by setting $s_x(0)=s_y(0)=s_z(0)=0$, for the
first qubit and $r_x(0)=r_y(0)=r_z(0)=0$ for the second qubit,
$Q_{ij}(0)=0$ for $i\neq j$ and $Q_{xx}(0)=Q_{yy}(0)=Q_{zz}(0)=-1$
for $\ket{\psi^{+}}$,  and $Q_{xx}(0)=Q_{yy}(0)=Q_{zz}(0)=1$ for
$\ket{\phi^{+}}$, $Q_{xx}(0)=Q_{zz}(0)=1, Q_{yy}(0)=-1$ for
$\ket{\psi^{-}}$ and so on. On the other hand, the users can use
an initial pure partial entangled state. This class of states is
characteristic by one parameter \cite{NM}. It is  defined by :
\begin{equation}\label{partIn}
\row{S}=(0,0,p), \quad \row{R}=(0,0,-p)~\mbox{and }
Q_{ij}=0 ~\mbox{for} ~i\neq j, Q_{xx}=Q_{yy}=-\sqrt{1-p^2} ~\mbox{and}
Q_{zz}=-1.
\end{equation}

\section{Quantum coding}
Let us consider that the partners Alice and Bob share  maximum or
partial entangled state. The aim of Alice is sending  the coded
information to Bob. But for some reasons the carrier of these
coded information is forced to pass through the Bloch channel. We
quantify the amount of information which decoded by Bob and
investigating the effect of the channel parameters and the type of
the initial carrier on the accuracy of the decoded information. To
show our idea, we implement the original dense coding protocol
which has been proposed by Bennett and Wienser \cite{Ben}.  This
protocol is  described as follows:
\begin{enumerate}
\item
Alice encodes two classical bits by using one of local unitary
operators.
\item
If Alice applies these unitary operators randomly
with probability $\eta_i$, then she codes the information in the
state,
\begin{equation}
\rho_{c}=\sum_{j=0}^{3}\Bigl\{\eta_j\mathcal{U}_j\otimes
I_2\rho^{out}_j\mathcal{U}_j^{\dagger}\otimes I_2\Bigr\},
\end{equation}
where $\rho^{out}$ is given by (\ref{out}) and $\mathcal{U}_j= I_1,
\sigma_{1x},\sigma_{1y},\sigma_{1y}$,
 are the unitary operators for
the first qubit and $I_2$ is the idenity operator for the second
qubit.
\item Alice sends her qubit to Bob, who makes  joint measurements on
the two qubits. The maximum amount of information which Bob can
extract from Alice's message is Bounded by,
\begin{equation}
I_{d}=\mathcal{S}\Bigl(\sum_{j=0}^{j=3}\eta_j\rho^{out}_j\Bigr)-\sum_{j=0}^{j=3}\eta_j\mathcal{S}(\rho^{out}_j).
\end{equation}
\end{enumerate}

\begin{figure}[t!]
  \begin{center}
 \includegraphics[width=19pc,height=15pc]{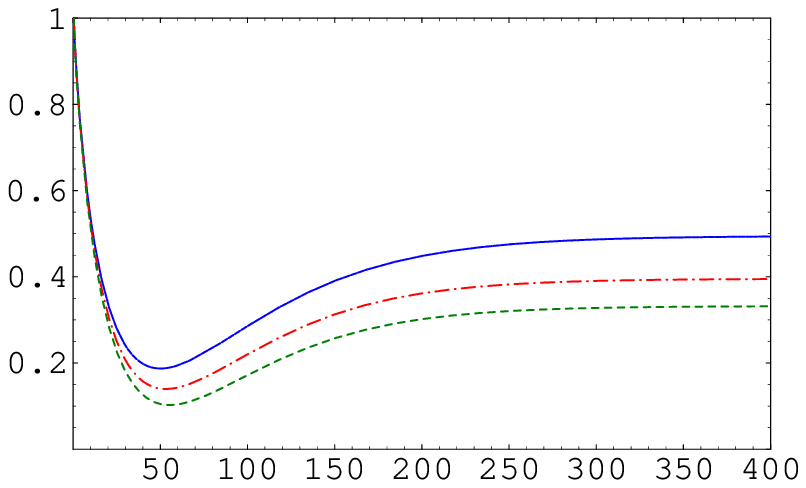}
 \includegraphics[width=19pc,height=15pc]{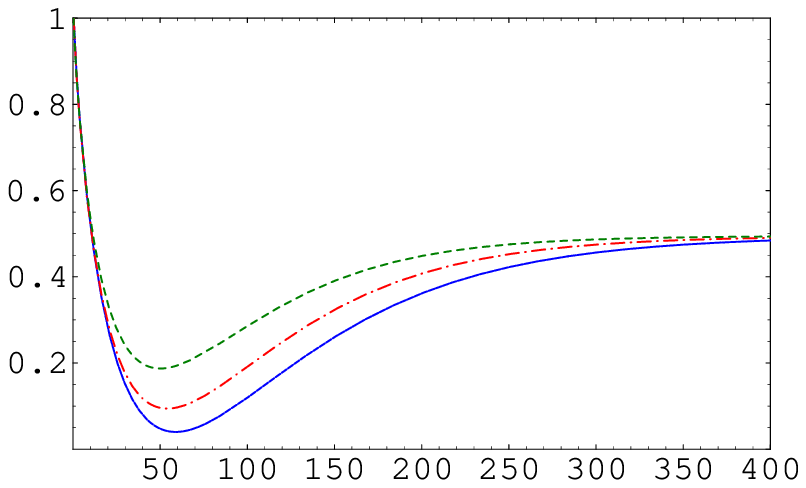}
\put(-470,90){$I_d$}
 \put(-232,90){$I_d$}
 \put(-257,155){$(a)$}
 \put(-25,155){$(b)$}
 \put(-340,-10){time}
 \put(-110,-10){time}
   \caption{The dynamics of the decoded information in a state initially prepared in maximum entangled state
    (a) the solid, dash-dot and dot curves are for
   $\expect{\sigma_{1z}}_{eq}=1,0.9,0.8$ respectively and $\expect{\sigma_{2z}}_{eq}=0.9,\alpha=0.5$.
   (b) the solid, dash-dot and dot lines for $\alpha=0.7,0.6,0.5$
   respectively and $\expect{\sigma_{1z}}_{eq}=1, \expect{\sigma_{2z}}_{eq}=0.9$.}
  \end{center}
\end{figure}

\begin{figure}
  \begin{center}
 \includegraphics[width=19pc,height=15pc]{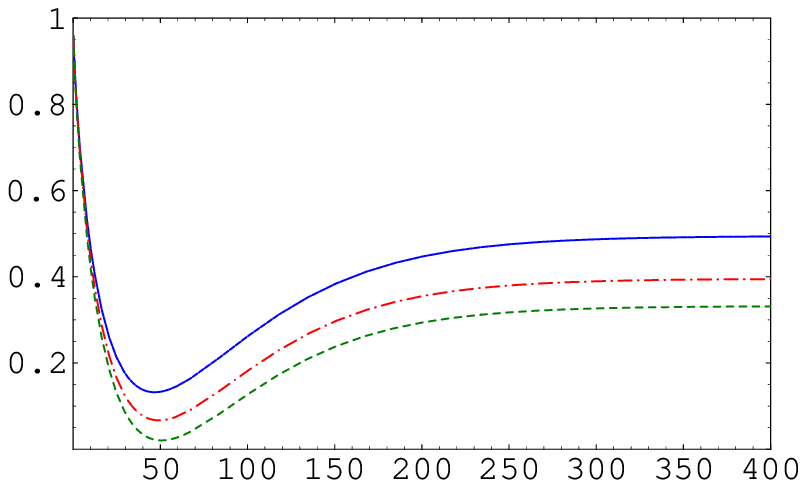}
 \includegraphics[width=19pc,height=15pc]{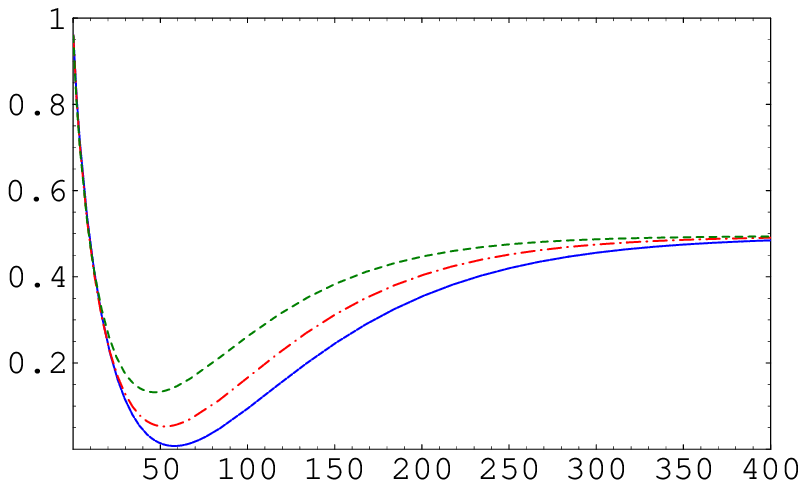}
 \put(-470,90){$I_d$}
 \put(-232,90){$I_d$}
 \put(-257,155){$(a)$}
 \put(-25,155){$(b)$}
 \put(-340,-10){time}
 \put(-110,-10){time}
   \caption{The same as Fig.(1), but for the partial entangled state.}
  \end{center}
\end{figure}

 Fig.(1a) displays the effect of the
equilibrium absolute values of the first qubit,
$\expect{\sigma_{1z}}_{eq}$ on the decoded information. It is
clear that, before the interaction is switched on, the decoded
information is very large and and goes down quickly once the
interaction is devolped. This decay of the decoded information
increases as $\expect{\sigma_{1z}}_{eq}$ decreases. However as
time increases, the decoded information increases gradually and
reaches its  upper bound.

The effect of the ratio between longitudinal and transverse
relaxation times, $\alpha_i$ is depicted in Fig.(1b). As
$\alpha_i$  increases and the equilibrium absolute values of the
two qubits are large, the decay of the decoded information becomes
faster. For $t>50$, the decoded information increases faster for
small values of $\alpha_i$. In Fig.(2), we consider that Alice
coded her information in partial  entangled state, where we set $p=0.5$ in (\ref{partIn}). The dynamics
of the decoded information is  similar to  that shown in Fig.(1).
However the maximum amount of the decoded information for the PES
is smaller than that for MES.

From Figs.(1 $\&$ 2), it is clear that the travelling coded
information in a state prepared initially in maximum entangled
states, MES is much better than using partial entangled states,
PES. This means that MES are more robust than PES for this type of
channels.

\section{Dynamics of Local and non Local Information}
Suppose we have a source supplies each user with a qubit
 to code their own  information. In this case, one says  that these
information are  local information. If  the qubits are forced to
pass through environment (say Bloch channels), then the two qubits
will entangled with each other and interact with the environment.
As a resultant of this interaction the local information will be
transferred between the two qubits and called non-local
information.

 In this section, we  investigate  the dynamics of  local
information $I_A$ which coded in Alice's qubit
$\rho_a=tr_b\{\rho_{c}\}$, $I_B$ is the local information which is
coded in Bob's qubit, $\rho_b=tr_a\{\rho_{c}\}$ and the non-local
information between Alice and Bob, $I_{ab}$ which is coded in the
state $\rho_{c}$ \cite{Metwally}. Due to the undesirable
interactions there are some information lose. These interactions
can be considered as another person (Eve), who tries to distill
information from the travelling state between  Alice to Bob.
Mathematically, this information is defined as:
\begin{equation}
I_{AE}=\mathcal{F}log\{\mathcal{F}\}+(1-\mathcal{F})log(1-\mathcal{F}),
\end{equation}
where $\mathcal{F}$ is the fidelity that Bob decoded the
information.

\begin{figure}[b!]
  \begin{center}
 \includegraphics[width=19pc,height=15pc]{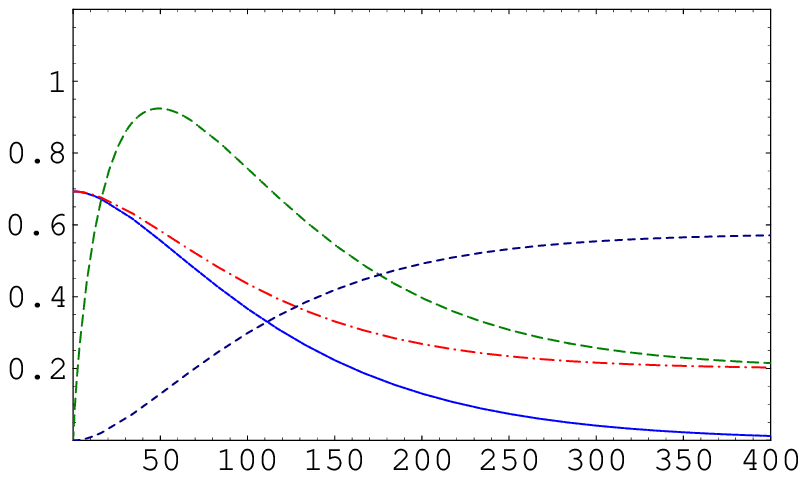}
  \includegraphics[width=19pc,height=15pc]{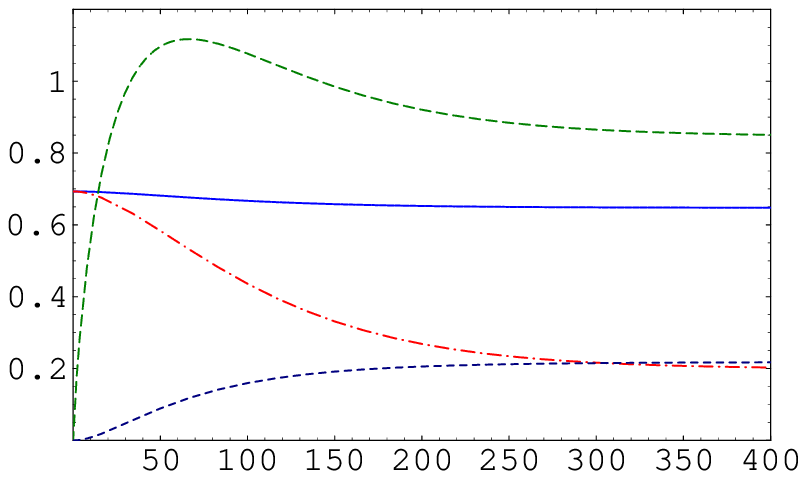}
\put(-257,155){$(a)$}
 \put(-25,155){$(b)$}
 \put(-340,-10){time}
 \put(-110,-10){time}
    \caption{The dynamics of the local and non-local information
   for $\alpha=0.7,\expect{\sigma_{2z}}_{eq}=0.9$. The solid,
   dash-dot,long-dash and dot curves for $I_A,I_B, I_{AB}$ and
   $I_{AE}$ respectively.
   (a)$\expect{\sigma_{1z}}_{eq}=1(b)\expect{\sigma_{1z}}_{eq}=0.3$.}
  \end{center}
\end{figure}

Figs.(3 $\&$ 4), describe the  effect of the channel's parameters on
the dynamics of the local information $I_A, I_B$, and  the
non-local information between Alice and Bob, $I_{AB}$, and between
Alice and Eve, $I_{AE}$. In Fig.(3), we investigate the effect of
the absolute equilibrium values on the dynamics of the travelling
coded information where we assume that Alice has coded her
information in  maximum entangled state. It is clear that, at the
beginning the  non-local information between Alice and Bob,
$I_{AB}$ and between Alice and Eve, $I_{AE}$ are zero, while $I_A$
and $I_B$ are non-zero. As soon as the interaction times  goes on,
$I_{AB}$ and $I_{AE}$ increase on the expanse of the local
information owned by Alice and Bob. As  time increases more, Eve
distill more information from Alice  and $I_{AE}$ is much larger
than $I_{A}$. On the other hand, due to the lose of  the
information from Alice side, the non-local information between
Alice and Bob decreases. The dynamical behavior of these different
types of the information are  depicted in Fig.(3a). Fig.(3b)
describes the dynamics of the local and non-local information for
small value of $\expect{\sigma_{1z}}_{eq}$ (say$\simeq 0.3)$. In
this case, the amount of information which is distilled by Eve  is
smaller than that shown in Fig.(3a). However, Alice's information
is slightly affected. Therefore  decreasing the absolute
equilibrium value of one qubit,  maximize the
non-local-information between the two qubits.

Fig.(4) displays the dynamics of  information, where Alice has coded
her information in partial entangled state. In this case, for
large values of $\expect{\sigma_{1z}}_{eq}$ and
$\expect{\sigma_{2z}}_{eq}$ the information which is gained by
Eve, increases abruptly on the expanse of Alice's
information and for $t>100$, $I_{AE}>I_{AB}$. However as one of
the absolute equilibrium values is decreased, the non-local
information between Alice and Bob, $I_{AB}$ is increased  very
fast and its maximum value is always larger than that depicted in
Fig.(3). As time goes on, $I_{AB}$, decreases slowly and its
minimum value is always larger than that depicted in Fig.(3b).
Although Eve's information increases fast, but $I_{AE}<I_{AB}$. In
a very small range of time $I_{AE}>I_{A}$. So for this choice of
the channel's parameters, Alice and Bob can communicate safely for
long range of time.
\begin{figure}[b!]
  \begin{center}
 \includegraphics[width=19pc,height=15pc]{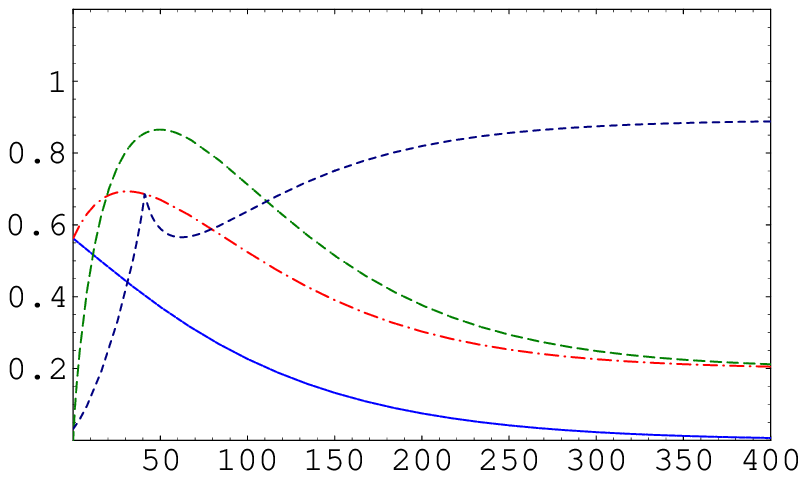}
 \includegraphics[width=19pc,height=15pc]{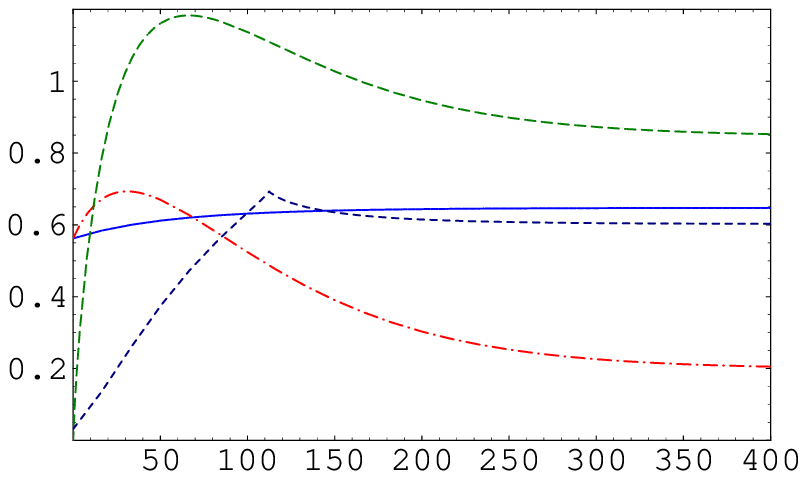}
\put(-257,155){$(a)$}
 \put(-25,155){$(b)$}
 \put(-340,-10){time}
 \put(-110,-10){time}
   \caption{The same as Fig.(3) but for the partial entangled
   state.}
  \end{center}
\end{figure}

The effect  of the ratio between longitudinal  and transverse
relaxation times, $\alpha_i$ is shown in Fig.(5), where we set
$\alpha_1=\alpha_2=0.3$, $\expect{\sigma_{1z}}_{eq}=1$ and
$\expect{\sigma_{2z}}_{eq}=0.9$. It is clear that, from Fig.(5a),
( we assume that Alice coded her information in MES), $I_{A}$
decreases very fast and Alice's state  turns into a completely
mixed state for $t>200$ and consequently Eve's information
increases very fast and reaches its maximum value faster than that
shown in Fig.(4a), in which  $\alpha_i=0.7$. As soon as Alice
loses her information completely, $I_{AB}$ and $I_{B}$ have
asymptotically the same values  $t>200$, which is much earlier
than that displayed in Fig.(3a). In Fig.(5b), we assume that the
information is initially coded in PES. In general, the dynamics of
information is similar to that shown in Fig.(5a), but from
Fig.(4a) and Fig.(5b), we can see that the safely communicate time
decreases and the non-local information between Alice and Bob,
$I_{AB}$ decreases very fast.

From Figs.(4 $\&$ 5), one concludes that the absolute equilibrium
values and the ratio between longitudinal  and transverse
relaxation times can be considered  as control parameters.  One
can improve the local information for one qubit by decreasing the
equilibrium values of the other qubit. Also, the eavesdropper
information can be minimized be decreasing the absolute
equilibrium of one qubit and increasing  the ratio between
longitudinal and transverse relaxation times. In this case, the
information lose is always smaller than the information between
the sender and receiver. Therefore, the users can increase the
safety communication time and  improve the non-local information.
\begin{figure}
  \begin{center}
 \includegraphics[width=19pc,height=15pc]{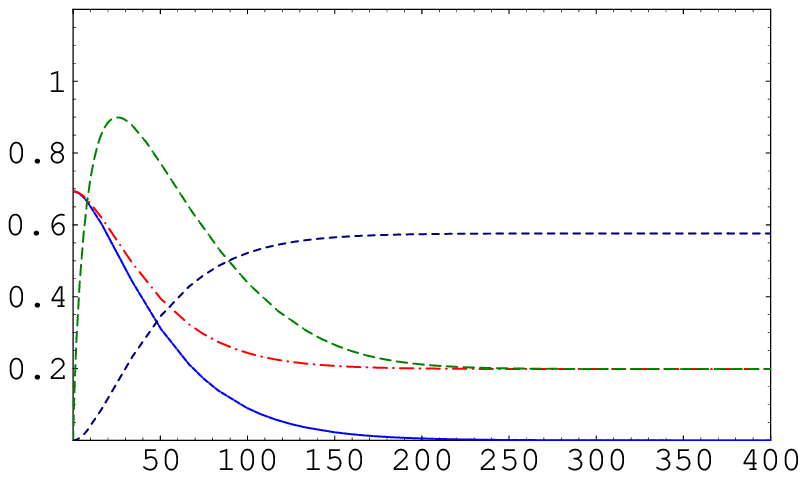}
 \includegraphics[width=19pc,height=15pc]{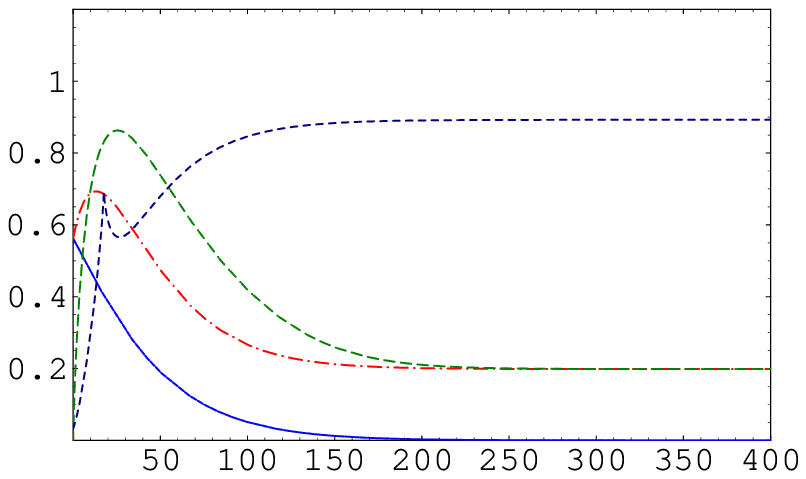}
 \put(-257,155){$(a)$}
 \put(-25,155){$(b)$}
 \put(-340,-10){time}
 \put(-110,-10){time}
   \caption{$\expect{\sigma_{1z}}_{eq}=0.9$ and $\alpha=0.3$ (a) for the maximum entangled state
   (b) For the partial entangled state.}
  \end{center}
\end{figure}

Also, the initial state which is used to code information plays an
important role on the secure communication. It is clear that,
coding information in maximum  entangled state is much better than
using partial entangled states, where for the first the users can
increase the safe time of communication by controlling on the
channel parameters.

\section{Conclusion}

The time evaluation of  a system consists of two qubits passes
through Bloch channel is investigated. The  quantum dense coding
protocol  is implemented by using two different initial states
setting:maximum and partial entangled states. The coded
information  is send with high accuracy by increasing the  absolute
equilibrium values of the two qubits and decreasing  the ratio of
the longitudinal  and transverse relaxation times. However, if the
absolute equilibrium value of one qubit decreases, the decoded
information decreases. It is shown that, using   maximum entangled
state for coding information is much better than using partial
entangled state. This means that, the maximum entangled states are
more robust then partial entangled states when they travel through
Bloch channels.

The local and non-local information are quantified for different
values of the channel parameters. There are some cases, where the
eavesdropper can distill more information on the expanse of the
travelling coded information. However the partners can communicate
safely when the non-local information between the two users is
larger than  that distilled from the travelling coded information.
Also, the absolute equilibrium values and the ratio of the
longitudinal  and transverse relaxation times can be considered as
a control parameters. It is clear that, for large values of the
equilibrium absolute parameters  for both qubit, the local
information of both qubit decreases faster  and consequently the
information gained by eavesdropper increases. However, if the
equilibrium absolute value of one qubit decreases, its
corresponding local information is slightly affected. Therefore,
to send the coded information from the sender to the receiver
safely, one has to decrease the absolute equilibrium value. Also,
as one increases the ratio of the longitudinal  and transverse
relaxation times, the survival time of the local and non-local
information increases.

\end{document}